\documentclass{article}

    \PassOptionsToPackage{numbers}{natbib}
    \usepackage[preprint]{neurips_2021}

\usepackage[utf8]{inputenc} % allow utf-8 input
\usepackage[T1]{fontenc}    % use 8-bit T1 fonts
\usepackage{hyperref}       % hyperlinks
\usepackage{url}            % simple URL typesetting
\usepackage{booktabs}       % professional-quality tables
\usepackage{amsfonts}       % blackboard math symbols
\usepackage{nicefrac}       % compact symbols for 1/2, etc.
\usepackage{microtype}      % microtypography
\usepackage{xcolor}         % colors
\usepackage{graphicx}
\title{DEM Super-Resolution with EfficientNetV2}

\author{%
   Bekir Z Demiray \\
   IIHR — Hydroscience \& Engineering \\
   University of Iowa \\
   Iowa City, IA \\
   \texttt{bekirzahit-demiray@uiowa.edu} \\
   \And
  Muhammed Sit \\
  IIHR — Hydroscience \& Engineering\\
  University of Iowa\\
  Iowa City, IA \\
  \texttt{muhammed-sit@uiowa.edu} \\
   \And
   Ibrahim Demir \\
   IIHR — Hydroscience \& Engineering \\
   University of Iowa \\
   Iowa City, IA \\
   \texttt{ibrahim-demir@uiowa.edu}
}

\begin{document}

\maketitle

\begin{abstract}
    Efficient climate change monitoring and modeling rely on high-quality geospatial and environmental datasets. Due to limitations in technical capabilities or resources, the acquisition of high-quality data for many environmental disciplines is costly. Digital Elevation Model (DEM) datasets are such examples whereas their low-resolution versions are widely available, high-resolution ones are scarce. In an effort to rectify this problem, we propose and assess an EfficientNetV2 based model. The proposed model increases the spatial resolution of DEMs up to 16 times without additional information.
\end{abstract}

\section{Introduction}
Digital Elevation Models (DEMs) are elevation-based representations of terrain surfaces. In the environment and climate change literature, DEMs have been used in variety of research problems over the years. That includes but is not limited to flood risk and hazard mapping \cite{1}, stream network extraction \cite{2}, understanding urban characteristics \cite{3}, surface texture analysis \cite{4}, and drainage network development \cite{5}. Furthermore, in flood mapping, using higher resolution DEMs yields better results in terms of accuracy of the constructed flood map when compared to ones sourced from low-resolution DEMs \cite{6}.

Water resource management and hydrological modeling using physically-based or data-driven approaches \cite{7} need not only DEMs but high-resolution DEMs for accurate hydrological predictions \cite{8}. Web-based tools for efficient disaster management, recovery, and response such as decision support \cite{9}, and data analytics \cite{10} systems that share climate change-related communication rely on high-quality DEM. Consequently, more efficient hydrological modeling for better climate change monitoring and modeling would highly benefit from higher resolution DEMs.

LiDAR (light detection and ranging) is the predominant way of generating high-resolution DEMs. LiDAR is an optical remote-sensing technique that estimates the distance between the sensor and the object, as well as the object's reflected energy. Even though it has been widely utilized and commonplace, the DEM production using LiDAR is still resource-intensive in terms of cost, time and computing power. The cost of the process increases drastically as the resolution of the product increases. Consequently, LiDAR products are often satellite based and low resolution, and as high-resolution ones are costly and harder to produce, they remain scarce. One way to increase the resolution of existing low-resolution DEMs produced by LiDAR without substantially increasing cost is to utilize data-driven approaches.

\subsection{Related Work}

Literature on increasing the DEM resolution with deep learning is limited. \cite{11} provides a proof-of-concept super-resolution approach for DEMs using convolutional neural networks. \cite{12} presents a new approach taking advantage of the state-of-the-art computer vision literature using residual blocks. Furthermore, \cite{13} explorers the power of GANs to increase the resolution up to 16x times.

Being a 2D tensor, a DEM is a data structure that many approaches in computer vision could be applied on \cite{11, 12, 14}. Thus, the literature on image super-resolution is of importance for this study. There are many studies that employ various neural network architectures for single image super-resolution. \cite{15} sets a milestone in super-resolution of images using CNNs while \cite{16, 17, 18} advances the effort to better accuracies. In another attempt, \cite{19} proposes using Generative Adversarial Networks (GANs) in image super-resolution and many others \cite{20, 21} alter the approach with different network architectures for both generator and discriminator for more efficiency. 

The recent studies on computer vision including super-resolution not only focus on the performance of the models, but they aim to create more efficient networks. Various manuscripts developed new methods and lightweight architectures such as post-upsampling or recursive learning \cite{22}. In accordance with these efforts, EfficientNet was introduced to find the balance between the network’s depth, width, and resolution to achieve better performance with smaller networks and it provided strong results on ImageNet and CIFAR/Flower/Cars \cite{23, 24} datasets.

To provide a step towards better climate change monitoring and modeling, in this paper, the power of EfficientNetV2 \cite{24} is harnessed to create a deep neural network model that aims to convert low-resolution DEMs into high-resolution DEMs up to 16x higher resolution without the need for additional data. Utilization of EfficientNetV2 was done by changing the core element of it, MobileNets \cite{26}. MobileNet is developed and improved by Google for mobile and embedded vision applications such as object detection or fine-grain classification \cite{25, 26, 27}. MobileNet introduced depth-wise separable convolutions to replace traditional convolutions \cite{25}. Beyond that, MobileNetV2 used the linear bottleneck and inverted residual architecture to provide a better efficient network than the previous model \cite{26}. MobileNetV3 combines MobileNetV2 with the squeeze and excite unit in the residual layers with different activation functions \cite{27}. To determine the efficiency of the proposed EfficientNetV2 based network, its performance is compared to that of classic interpolation methods such as bicubic and bilinear alongside artificial neural network-based models, and preliminary results are shared.

The rest of this paper is structured as follows; in section 2, the dataset used in this work is introduced and the overview of the methodology is presented. Then in section 3, the preliminary results for compared methods and the proposed neural network are presented and discussed. Finally, in section 4, we summarize the outcomes.

\begin{table}[h]
  \caption{Statistical Summary of Dataset}
  \label{dataset-table}
  \centering
  \begin{tabular}{llll}
    \toprule
    & Avg. Elevation (m)     & Min. Elevation (m)    & Max. Elevation (m) \\
    \midrule
    Training &  653.1  &  205.7 & 984.9    \\
    Test     &  621.7  &  230.0  & 982.7 \\
    \bottomrule
  \end{tabular}
\end{table}

\section{Method}
\subsection{Data}
The data utilized in this work was acquired from the North Carolina Floodplain Mapping Program. It is a state program that allows the public to download various data types such as bare earth DEMs or flood zones for North Carolina through the web interface. The dataset we collected for this study covers a total area of 732 km2 from two different counties (i.e., Wake and Guilford) in North Carolina, United States. A total area of 590 km2 is used as the training set, and the remainder of the dataset, which covers a total area of 142 km2, as the test set. Each of the used DEMs was collected at a spacing of approximately 2 points per square meter. In the experiments, DEMs with a resolution of 3 feet and 50 feet are used as high-resolution and low-resolution examples, respectively. The North Carolina Floodplain Mapping Program supplied the data as tiles. Each of the high-resolution DEM tiles is delivered as 1600x1600 data points and the low-resolution DEM tiles as 100x100 data points. In the preprocessing phase, high-resolution DEMs were split into 400×400 data points and low-resolution DEMs were split into 25×25 data points to decrease the computational need. In addition to the fragmentation process, DEMs with missing values are discarded from the dataset prior to the training. Table \ref{dataset-table} shows the average, minimum and maximum elevation values for both datasets.

\subsection{Neural Network}
\begin{figure}[h]
  \centering
  \includegraphics[scale=0.4]{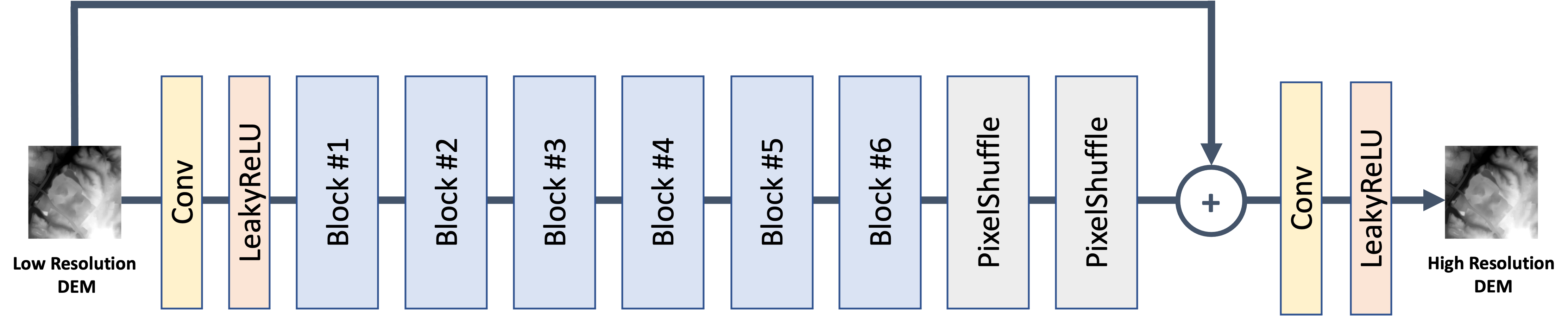}
  \caption{Architecture of Proposed Method}
  \label{network-pic}
\end{figure}

As mentioned before, the base for this work is EfficientNetV2. In order to create a model that carries out the DEM super-resolution task that is converting a 25x25 tensor into a 400x400 one, we alter EfficientNetV2 by modifying its core component, MobileNets, which was not designed with a super-resolution task in mind. Our model takes a 50 feet DEM as low-resolution data and passes it to a convolution layer with 24 feature maps prior to feeding it to multiple MobileNetV3 blocks. After that, two sub-pixel convolutional layers \cite{28} are used to upscale to reach desired high resolution product. In the end, the output of upsampling blocks is summed up with an interpolated version of low-resolution input data, then the summation is fed to a final convolutional layer which in turn gives the output of the network, a 3 feet high-resolution DEM. The visual representation of our model is provided in Figure \ref{network-pic}.

In the proposed model, we modify the MobileNetV3 blocks to use for super-resolution tasks. Similar to prior studies in image super-resolution super-resolution \cite{21, 29}, batch normalization layers were removed from MobileNetV3 blocks in accordance with the experimental results on model development. In addition, h-swish activation layers were replaced with LeakyReLU activation functions. Contrary to MobileNetV3 and EfficientNetV2, in our model, input resolutions are not changed until the upsampling blocks, similarly to other super-resolution models \cite{19, 21}. The remaining parts of the MobileNetV3 are the same as the original implementation including SE blocks. For more information about MobileNetV3’s implementation details, please refer to \cite{24, 27}. For the cardinality and design of the MobileNetV3 blocks in this work, we followed the same network design with EfficientNetV2-S architecture as shown in Table \ref{block-table}. 

\begin{table}[h]
  \caption{MobileNet blocks in our model}
  \label{block-table}
  \centering
  \begin{tabular}{lllllll}
    \toprule
    & Block\#1 & Block\#2 & Block\#3 & Block\#4 & Block\#5 & Block\#6 \\
    \midrule
    \# of Channels  &  24   &  48 &  64 &  128 &  160 &  256    \\
    Expansion ratio &  1 &  4 &  4 &  4   &  6  & 6    \\
    \# of Layers &  2 &  4 &  4 &  6   &  9  & 15    \\
    \bottomrule
  \end{tabular}
\end{table}

For training, Adam \cite{30} was used as the optimizer with a learning rate of 0.001, and the MSE loss was used as the cost function. The network is implemented with Pytorch 1.9 \cite{31} and ReduceLROnPlateau from Pytorch is used as the scheduler to reduce the learning rate when there are no improvements on model loss for 10 consecutive epochs. The training was done on NVIDIA Titan V GPUs.

\section{Results}
It is a typical practice to employ MSE to evaluate the performance of proposed approaches in DEM super-resolution \cite{11, 14, 32}. Considering that DEM provides height values for the corresponding area, it is reasonable to utilize a metric such as MSE that is representative of quantitative measurements in order to better understand how well a method performs. Consequently, we used MSE scores to compare our methods with various approaches. Classical approaches such as bicubic and bilinear interpolations are widely used to show the comparable performance of proposed DEM super-resolution methodologies and we also report their performance over our dataset to form a baseline. Bicubic and bilinear interpolation implementations in scikit-image library \cite{33} were employed for this purpose. In addition to classical approaches, we compared our results with three deep learning studies. D-SRCNN \cite{11} is a CNN-based method for increasing the resolution of a DEM with a similar architecture to SRCNN \cite{15}. DPGN \cite{12} is also a CNN-based model that uses skip-connections and residual blocks to increase DEM resolution. Lastly, D-SRGAN \cite{13}, which is a generative adversarial network that aims to increase the resolution up to 4x scale factors. Table \ref{result-table} shows the mean squared errors of different methods on both training and test datasets for elevation data. According to experiment results, all neural networks outperform the classical methods, but our method, noted as EfficientNetV2-DEM, gives the best results with a significant margin on both datasets.

\begin{table}
  \caption{The Performance Comparison of Different Methods as MSE in meters}
  \label{result-table}
  \centering
  \begin{tabular}{lllllll}
    \toprule
         &Bicubic &Bilinear &D-SRCNN &DPGN &D-SRGAN &EfficientNetV2-DEM \\
    \midrule
    Training  &  0.968  &  1.141 &  0.900 &  0.758 & 0.766 &  0.625 \\
    Test  &  0.946 &  1.124 &  0.872 &  0.803 &  0.753  &  0.640 \\
    \bottomrule
  \end{tabular}
\end{table}

\begin{figure}[h]
  \centering
  \includegraphics[scale=0.4]{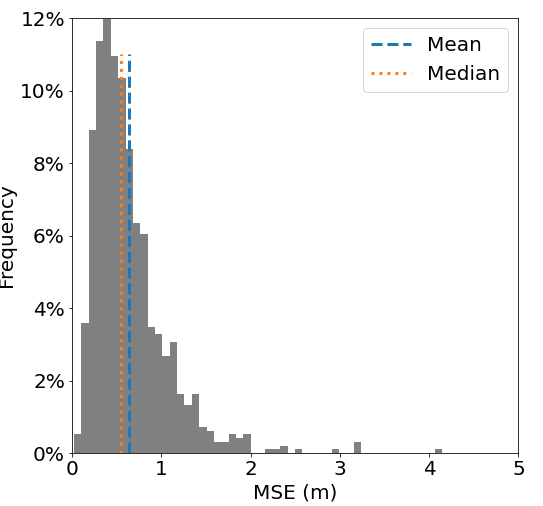}
  \caption{Error Distribution of EfficientNetV2-DEM Model on Test Dataset}
  \label{errorpic}
\end{figure}

The error distribution of the proposed network on testing data is also provided in Figure \ref{errorpic}. The mean, median, and standard deviation of errors in testing data for EfficientNetV2-DEM are 0.64, 0.54, and 0.42 m, respectively. According to experiment results, \%76 and \%80 of all predicted values are within plus or minus one standard deviation from the mean on training and test dataset, respectively, which shows that the performance of the model we propose is robust with a limited number of outliers.

\section{Conclusions}
In this study, we presented an EfficientNetV2 based DEM super-resolution model that increases the spatial resolution of DEMs up to 16 times without the need for any additional information. The efficacy of the proposed network is shown by comparing its performance with conventional interpolation methods and artificial neural network-based studies. As it was briefly discussed in the Results section over the preliminary results, the proposed network shows promise by surpassing the compared methods by a significant margin. Considering the EfficientNetV2 based network’s training times are manageable, we believe it provides a comparable alternative to both classical and machine learning-based methods.

As future directions for this approach, since we provided the proof of concept here, we aim to build a larger network to train it with larger DEM datasets. To create a neural network that grasps the correlation between low-resolution and high-resolution DEMs better, another future aspect is to create a custom cost function instead of using MSE.

As addressing climate change needs better datasets \cite{34, 35}, we believe this study, along with its future perspectives, provides a great starting point in an effort to improve environmental datasets.

\bibliography{main}

% Generated by IEEEtranN.bst, version: 1.14 (2015/08/26)
\begin{thebibliography}{35}
\providecommand{\natexlab}[1]{#1}
\providecommand{\url}[1]{#1}
\csname url@samestyle\endcsname
\providecommand{\newblock}{\relax}
\providecommand{\bibinfo}[2]{#2}
\providecommand{\BIBentrySTDinterwordspacing}{\spaceskip=0pt\relax}
\providecommand{\BIBentryALTinterwordstretchfactor}{4}
\providecommand{\BIBentryALTinterwordspacing}{\spaceskip=\fontdimen2\font plus
\BIBentryALTinterwordstretchfactor\fontdimen3\font minus
  \fontdimen4\font\relax}
\providecommand{\BIBforeignlanguage}[2]{{%
\expandafter\ifx\csname l@#1\endcsname\relax
\typeout{** WARNING: IEEEtranN.bst: No hyphenation pattern has been}%
\typeout{** loaded for the language `#1'. Using the pattern for}%
\typeout{** the default language instead.}%
\else
\language=\csname l@#1\endcsname
\fi
#2}}
\providecommand{\BIBdecl}{\relax}
\BIBdecl

\bibitem[Li and Wong(2010)]{1}
J.~Li and D.~W. Wong, ``Effects of dem sources on hydrologic applications,''
  \emph{Computers, Environment and urban systems}, vol.~34, no.~3, pp.
  251--261, 2010.

\bibitem[Tarboton(1997)]{2}
D.~G. Tarboton, ``A new method for the determination of flow directions and
  upslope areas in grid digital elevation models,'' \emph{Water resources
  research}, vol.~33, no.~2, pp. 309--319, 1997.

\bibitem[Priestnall et~al.(2000)Priestnall, Jaafar, and Duncan]{3}
G.~Priestnall, J.~Jaafar, and A.~Duncan, ``Extracting urban features from lidar
  digital surface models,'' \emph{Computers, Environment and Urban Systems},
  vol.~24, no.~2, pp. 65--78, 2000.

\bibitem[Trevisani et~al.(2012)Trevisani, Cavalli, and Marchi]{4}
S.~Trevisani, M.~Cavalli, and L.~Marchi, ``Surface texture analysis of a
  high-resolution dtm: Interpreting an alpine basin,'' \emph{Geomorphology},
  vol. 161, pp. 26--39, 2012.

\bibitem[Fairfield and Leymarie(1991)]{5}
J.~Fairfield and P.~Leymarie, ``Drainage networks from grid digital elevation
  models,'' \emph{Water resources research}, vol.~27, no.~5, pp. 709--717,
  1991.

\bibitem[Kim et~al.(2019)Kim, Gourbesville, and Liong]{6}
D.-E. Kim, P.~Gourbesville, and S.-Y. Liong, ``Overcoming data scarcity in
  flood hazard assessment using remote sensing and artificial neural network,''
  \emph{Smart Water}, vol.~4, no.~1, pp. 1--15, 2019.

\bibitem[Sit et~al.(2021)Sit, Demiray, and Demir]{7}
M.~Sit, B.~Demiray, and I.~Demir, ``Short-term hourly streamflow prediction
  with graph convolutional gru networks,'' \emph{arXiv preprint
  arXiv:2107.07039}, 2021.

\bibitem[Vaze et~al.(2010)Vaze, Teng, and Spencer]{8}
J.~Vaze, J.~Teng, and G.~Spencer, ``Impact of dem accuracy and resolution on
  topographic indices,'' \emph{Environmental Modelling \& Software}, vol.~25,
  no.~10, pp. 1086--1098, 2010.

\bibitem[Sermet et~al.(2020)Sermet, Demir, and Muste]{9}
Y.~Sermet, I.~Demir, and M.~Muste, ``A serious gaming framework for decision
  support on hydrological hazards,'' \emph{Science of The Total Environment},
  vol. 728, p. 138895, 2020.

\bibitem[Sit et~al.(2019)Sit, Sermet, and Demir]{10}
M.~Sit, Y.~Sermet, and I.~Demir, ``Optimized watershed delineation library for
  server-side and client-side web applications,'' \emph{Open Geospatial Data,
  Software and Standards}, vol.~4, no.~1, pp. 1--10, 2019.

\bibitem[Chen et~al.(2016)Chen, Wang, Xu, et~al.]{11}
Z.~Chen, X.~Wang, Z.~Xu \emph{et~al.}, ``Convolutional neural network based dem
  super resolution.'' \emph{International Archives of the Photogrammetry,
  Remote Sensing \& Spatial Information Sciences}, vol.~41, 2016.

\bibitem[Xu et~al.(2019)Xu, Chen, Yi, Gui, Hou, and Ding]{12}
Z.~Xu, Z.~Chen, W.~Yi, Q.~Gui, W.~Hou, and M.~Ding, ``Deep gradient prior
  network for dem super-resolution: Transfer learning from image to dem,''
  \emph{ISPRS Journal of Photogrammetry and Remote Sensing}, vol. 150, pp.
  80--90, 2019.

\bibitem[Demiray et~al.(2021)Demiray, Sit, and Demir]{13}
B.~Z. Demiray, M.~Sit, and I.~Demir, ``D-srgan: Dem super-resolution with
  generative adversarial networks,'' \emph{SN Computer Science}, vol.~2, no.~1,
  pp. 1--11, 2021.

\bibitem[Xu et~al.(2015)Xu, Wang, Chen, Xiong, Ding, and Hou]{14}
Z.~Xu, X.~Wang, Z.~Chen, D.~Xiong, M.~Ding, and W.~Hou, ``Nonlocal similarity
  based dem super resolution,'' \emph{ISPRS Journal of Photogrammetry and
  Remote Sensing}, vol. 110, pp. 48--54, 2015.

\bibitem[Dong et~al.(2015)Dong, Loy, He, and Tang]{15}
C.~Dong, C.~C. Loy, K.~He, and X.~Tang, ``Image super-resolution using deep
  convolutional networks,'' \emph{IEEE transactions on pattern analysis and
  machine intelligence}, vol.~38, no.~2, pp. 295--307, 2015.

\bibitem[Dong et~al.(2016)Dong, Loy, and Tang]{16}
C.~Dong, C.~C. Loy, and X.~Tang, ``Accelerating the super-resolution
  convolutional neural network,'' in \emph{European conference on computer
  vision}.\hskip 1em plus 0.5em minus 0.4em\relax Springer, 2016, pp. 391--407.

\bibitem[Kim et~al.(2016{\natexlab{a}})Kim, Lee, and Lee]{17}
J.~Kim, J.~K. Lee, and K.~M. Lee, ``Accurate image super-resolution using very
  deep convolutional networks,'' in \emph{Proceedings of the IEEE conference on
  computer vision and pattern recognition}, 2016, pp. 1646--1654.

\bibitem[Kim et~al.(2016{\natexlab{b}})Kim, Lee, and Lee]{18}
------, ``Deeply-recursive convolutional network for image super-resolution,''
  in \emph{Proceedings of the IEEE conference on computer vision and pattern
  recognition}, 2016, pp. 1637--1645.

\bibitem[Ledig et~al.(2017)Ledig, Theis, Husz{\'a}r, Caballero, Cunningham,
  Acosta, Aitken, Tejani, Totz, Wang, et~al.]{19}
C.~Ledig, L.~Theis, F.~Husz{\'a}r, J.~Caballero, A.~Cunningham, A.~Acosta,
  A.~Aitken, A.~Tejani, J.~Totz, Z.~Wang \emph{et~al.}, ``Photo-realistic
  single image super-resolution using a generative adversarial network,'' in
  \emph{Proceedings of the IEEE conference on computer vision and pattern
  recognition}, 2017, pp. 4681--4690.

\bibitem[Wang et~al.(2018{\natexlab{a}})Wang, Perazzi, McWilliams,
  Sorkine-Hornung, Sorkine-Hornung, and Schroers]{20}
Y.~Wang, F.~Perazzi, B.~McWilliams, A.~Sorkine-Hornung, O.~Sorkine-Hornung, and
  C.~Schroers, ``A fully progressive approach to single-image
  super-resolution,'' in \emph{Proceedings of the IEEE conference on computer
  vision and pattern recognition workshops}, 2018, pp. 864--873.

\bibitem[Wang et~al.(2018{\natexlab{b}})Wang, Yu, Wu, Gu, Liu, Dong, Qiao, and
  Change~Loy]{21}
X.~Wang, K.~Yu, S.~Wu, J.~Gu, Y.~Liu, C.~Dong, Y.~Qiao, and C.~Change~Loy,
  ``Esrgan: Enhanced super-resolution generative adversarial networks,'' in
  \emph{Proceedings of the European conference on computer vision (ECCV)
  workshops}, 2018, pp. 0--0.

\bibitem[Wang et~al.(2020)Wang, Chen, and Hoi]{22}
Z.~Wang, J.~Chen, and S.~C. Hoi, ``Deep learning for image super-resolution: A
  survey,'' \emph{IEEE transactions on pattern analysis and machine
  intelligence}, 2020.

\bibitem[Tan and Le(2019)]{23}
M.~Tan and Q.~Le, ``Efficientnet: Rethinking model scaling for convolutional
  neural networks,'' in \emph{International Conference on Machine
  Learning}.\hskip 1em plus 0.5em minus 0.4em\relax PMLR, 2019, pp. 6105--6114.

\bibitem[Tan and Le(2021)]{24}
M.~Tan and Q.~V. Le, ``Efficientnetv2: Smaller models and faster training,''
  \emph{arXiv preprint arXiv:2104.00298}, 2021.

\bibitem[Sandler et~al.(2018)Sandler, Howard, Zhu, Zhmoginov, and Chen]{26}
M.~Sandler, A.~Howard, M.~Zhu, A.~Zhmoginov, and L.-C. Chen, ``Mobilenetv2:
  Inverted residuals and linear bottlenecks,'' in \emph{Proceedings of the IEEE
  conference on computer vision and pattern recognition}, 2018, pp. 4510--4520.

\bibitem[Howard et~al.(2017)Howard, Zhu, Chen, Kalenichenko, Wang, Weyand,
  Andreetto, and Adam]{25}
A.~G. Howard, M.~Zhu, B.~Chen, D.~Kalenichenko, W.~Wang, T.~Weyand,
  M.~Andreetto, and H.~Adam, ``Mobilenets: Efficient convolutional neural
  networks for mobile vision applications,'' \emph{arXiv preprint
  arXiv:1704.04861}, 2017.

\bibitem[Howard et~al.(2019)Howard, Sandler, Chu, Chen, Chen, Tan, Wang, Zhu,
  Pang, Vasudevan, et~al.]{27}
A.~Howard, M.~Sandler, G.~Chu, L.-C. Chen, B.~Chen, M.~Tan, W.~Wang, Y.~Zhu,
  R.~Pang, V.~Vasudevan \emph{et~al.}, ``Searching for mobilenetv3,'' in
  \emph{Proceedings of the IEEE/CVF International Conference on Computer
  Vision}, 2019, pp. 1314--1324.

\bibitem[Shi et~al.(2016)Shi, Caballero, Husz{\'a}r, Totz, Aitken, Bishop,
  Rueckert, and Wang]{28}
W.~Shi, J.~Caballero, F.~Husz{\'a}r, J.~Totz, A.~P. Aitken, R.~Bishop,
  D.~Rueckert, and Z.~Wang, ``Real-time single image and video super-resolution
  using an efficient sub-pixel convolutional neural network,'' in
  \emph{Proceedings of the IEEE conference on computer vision and pattern
  recognition}, 2016, pp. 1874--1883.

\bibitem[Lim et~al.(2017)Lim, Son, Kim, Nah, and Mu~Lee]{29}
B.~Lim, S.~Son, H.~Kim, S.~Nah, and K.~Mu~Lee, ``Enhanced deep residual
  networks for single image super-resolution,'' in \emph{Proceedings of the
  IEEE conference on computer vision and pattern recognition workshops}, 2017,
  pp. 136--144.

\bibitem[Kingma and Ba(2014)]{30}
D.~P. Kingma and J.~Ba, ``Adam: A method for stochastic optimization,''
  \emph{arXiv preprint arXiv:1412.6980}, 2014.

\bibitem[Paszke et~al.(2019)Paszke, Gross, Massa, Lerer, Bradbury, Chanan,
  Killeen, Lin, Gimelshein, Antiga, et~al.]{31}
A.~Paszke, S.~Gross, F.~Massa, A.~Lerer, J.~Bradbury, G.~Chanan, T.~Killeen,
  Z.~Lin, N.~Gimelshein, L.~Antiga \emph{et~al.}, ``Pytorch: An imperative
  style, high-performance deep learning library,'' \emph{Advances in neural
  information processing systems}, vol.~32, pp. 8026--8037, 2019.

\bibitem[Argudo et~al.(2018)Argudo, Chica, and Andujar]{32}
O.~Argudo, A.~Chica, and C.~Andujar, ``Terrain super-resolution through aerial
  imagery and fully convolutional networks,'' in \emph{Computer Graphics
  Forum}, vol.~37.\hskip 1em plus 0.5em minus 0.4em\relax Wiley Online Library,
  2018, pp. 101--110.

\bibitem[Van~der Walt et~al.(2014)Van~der Walt, Sch{\"o}nberger,
  Nunez-Iglesias, Boulogne, Warner, Yager, Gouillart, and Yu]{33}
S.~Van~der Walt, J.~L. Sch{\"o}nberger, J.~Nunez-Iglesias, F.~Boulogne, J.~D.
  Warner, N.~Yager, E.~Gouillart, and T.~Yu, ``scikit-image: image processing
  in python,'' \emph{PeerJ}, vol.~2, p. e453, 2014.

\bibitem[Rolnick et~al.(2019)Rolnick, Donti, Kaack, Kochanski, Lacoste,
  Sankaran, Ross, Milojevic-Dupont, Jaques, Waldman-Brown, et~al.]{34}
D.~Rolnick, P.~L. Donti, L.~H. Kaack, K.~Kochanski, A.~Lacoste, K.~Sankaran,
  A.~S. Ross, N.~Milojevic-Dupont, N.~Jaques, A.~Waldman-Brown \emph{et~al.},
  ``Tackling climate change with machine learning,'' \emph{arXiv preprint
  arXiv:1906.05433}, 2019.

\bibitem[Sit et~al.(2020)Sit, Demiray, Xiang, Ewing, Sermet, and Demir]{35}
M.~Sit, B.~Z. Demiray, Z.~Xiang, G.~J. Ewing, Y.~Sermet, and I.~Demir, ``A
  comprehensive review of deep learning applications in hydrology and water
  resources,'' \emph{Water Science and Technology}, vol.~82, no.~12, pp.
  2635--2670, 2020.

\end{thebibliography}

\end{document}